\documentclass[12pt,a4paper]{article}% ?? twoside ??

\usepackage[english]{babel}
\usepackage{amsmath,amssymb,amsfonts,amsthm}
\usepackage[mathscr]{eucal}
\usepackage[all]{xy}
\usepackage{hyperref}
\usepackage{setspace}
\usepackage{upgreek}
\usepackage{hyperref}%hypertex
\usepackage{cite}

\usepackage{times}
\usepackage{color}

\usepackage{indentfirst}

\allowdisplaybreaks

\begin{document}
\title{A stable higher-derivative theory with the Yang-Mills gauge symmetry}
\author {D.S.~Kaparulin}
\date{\small\textit{
Physics Faculty, Tomsk State University, Lenina ave. 36, Tomsk
634050, Russia}}

\maketitle

\begin{abstract}
\noindent An example of higher-derivative theory with a non-Abelian gauge symmetry is proposed.  In the free limit, the model describes 
the multiplet of vector fields, being subjected to the extended Chern-Simons equations. The theory admits a single second-rank conserved 
tensor, whose 00-component can be bounded or unbounded from below depending on the model parameters. If the conserved tensor 
has a bounded 00-component, the dynamics is stable. The equations of motion are non-Lagrangian.
\end{abstract}

\section{Introduction}

The stability of higher-derivative dynamics is studied for decades \cite{KLS14, Woodard, Pavsic, Smilga, Boulanger}. The majority of attention is attracted to the higher-derivative 
gravity models, where some particular examples of stable theories are known \cite{Tomboulis}. The dynamics of $f(R)$-gravity models is 
stabilised by the special structure of constants in the Hamiltonian formalism. The Yang-Mills theory is another important 
high energy physics model, which admits the higher-derivative generalisation \cite{SlavnovBook}. This model suffers from the Ostrogradsky instability already at the free level.
The inclusion of variational interactions does not stabilise the dynamics \cite{DaiJ-1}. This means that all the higher-derivative Lagrangian 
models with the Yang-Mills gauge symmetry are unstable. This conclusion does not applies to the interactions 
that do not come from the variational principle.

In the recent years, the extended Chern-Simons model \cite{Jackiw} is often considered as a lower-dimensional analog of 
the higher-derivative Yang-Mills. In the paper \cite{KKL15}, it has been observed that the 
extended Chern-Simons theory admits a series of bounded conserved quantities that stabilise the dynamics at the free level, even though the canonical energy 
of the model is unbounded from below. In the paper \cite{AKL18}, a class of consistent interactions between the charged scalar field has 
been proposed such that preserves a selected representative of conserved quantity series of the free model. The stable interactions are non-Lagrangian, but the dynamics admits the Hamiltonian form with a bounded Hamiltonian \cite{AKL19}. Up to date, the all the known stable interactions have Abelian gauge symmetry. Non-Abelian interactions are also known, but the dynamics is unstable already at free level \cite{DaiJ-2}.

In the current article, we address the issue of construction of stable interacting theory with a non-Abelian gauge symmetry. We show that the free 
extended Chern-Simons model of third order admits a consistent deformation of equations of motion preserving the gauge symmetries and gauge identities 
of free model. A single representative in the symmetric second-rank tensor series of free model is conserved at the non-linear level. 
Depending on the coupling parameters, its 00-component can be bounded or unbounded. The stable interaction vertices inevitably correspond to non-Lagrangian equations of motion. The consistency of interaction is verified by means of the involutive form of the equations \cite{KLS13}. It is explicitly shown that the model admits a gauge identity. 
The number of physical degrees of freedom appears to be the same as in the free theory.

The rest of the article is organised as follows. In Section 2, I describe the non-linear equations and examine the consistency of interaction. In Section 3, 
I present the second-rank conserved tensor, and examine the stability of the theory. 

\section{The model}
We consider the theory of vector field multiplet in three-dimensional Minkowski space with the local coordinates $x^\mu,\mu=0,1,2$. A mostly positive signature of the Minkowski metric is used throughout the paper. The dynamical variables are the vector fields $A{}^a{}_\mu(x)$, where $a=1,\ldots,n,$ is the isotopic index. The equations of motion are introduced in the following form:
\begin{equation}\label{EoM}\begin{array}{c}\displaystyle
\mathrm{T}{}^a{}_\mu=K{}^a{}_\mu+\alpha_2G{}^a{}_\mu+\alpha_1F{}^{a}{}_\mu+\frac{ig\gamma}{2} f{}^{abc}\varepsilon_{\mu\nu\rho}X^b{}^\nu X^c{}^\rho=0\,,\\[5mm]\displaystyle
X{}^a{}_\mu=G{}^a{}_\mu+\beta F{}^a{}_\mu\,,\qquad \gamma=(\alpha_1-\alpha_2\beta+\beta^2)^{-1}\,.
\end{array}\end{equation}
The vectors $K{}^a{}_\mu(x),G{}^a{}_\mu(x), F{}^a{}_\mu(x)$ denote the Yang-Mills field strength (the dual of the Yang-Mills strength tensor), 
and its higher-derivative generalisations,
\begin{equation}\begin{array}{c}\displaystyle
F_{\mu}{}^a=\varepsilon_{\mu\nu\rho}F^{a\nu\rho},\qquad F{}^a{}_{\nu\rho}=\partial_\nu A{}^{a}{}_{\rho}-\partial_\rho A{}^{a}{}_{\nu}+
igf{}^{abc}A{}^b{}_\nu A{}^c{}_\rho\,,\\[5mm]\displaystyle
G{}^a{}_\mu=\varepsilon_{\mu\nu\rho}(\partial^\nu F{}^a{}_\mu+igf{}^{abc}A{}^{b\nu} F{}^{c\rho})\,,\quad K{}^a{}_\mu(x)=\varepsilon_{\mu\nu\rho}(\partial^\nu G{}^a{}_\mu(x)+igf{}^{abc}A{}^{b\nu} G{}^{c\rho})\,.
\end{array}\end{equation}
The quantities $f{}^{abc}$ represent the structure constants of the Lie algebra of a semisimple Lie group. The constants $f^{abc}$ are supposed to be totally antisymmetric with 
respect to the transposition of indices, and they meet to the Jacobi identity,
\begin{equation}\label{Jid}
	f^{abc}f^{ade}+\text{cycle}(c,d,e)=0\,.
\end{equation}
The summation over the repeated Latin indices $a,b,c,\ldots$ is implied. The upper and lower colour indices are identified.
The quantity $\varepsilon_{\mu\nu\rho}$ stands for $3d$ Levi-Civita symbol. The convention $\varepsilon_{012}=1$ is used.
The covariant derivative is determined by the usual rule, 
\begin{equation}\label{D-cov}
(D_{\mu}){}^{ab}=\delta{}^{ab}\partial_\mu+igf^{abc}A{}^b{}_\mu\,,\qquad [D_\mu,D_\nu]{}^{ab}=ig\varepsilon_{\mu\nu\rho}f^{abc}F{}^a{}^\rho\,. 
\end{equation}
Relations (\ref{Jid}), (\ref{D-cov}) imply that the vectors $F{}^a{}_\mu, G{}^a{}_\mu$ are covariantly transverse,
\begin{equation}\label{FG-tr}
	D^\mu F_\mu=0\,,\qquad D^\mu G_\mu=0\,.
\end{equation}
The real numbers $\alpha_2$, $\alpha_1$, $\beta$, $g$ are model parameters. The constants $\alpha_1,\alpha_2$ determine the free limit of equations (\ref{EoM}).
The coupling constant is $g$. The parameter $\beta$ distinguishes the various interacting theories with one and the same value of the coupling constant $g$.

Equations (\ref{EoM}) represent a non-Ableian generalisation of the well-known extended Chern-Simons theory of third order \cite{Jackiw}. 
The equations of motion are preserved by the standard Yang-Mills gauge symmetry
\begin{equation}\label{gt}
\delta_\varepsilon A{}^a{}_\mu(x)=D_\mu\varepsilon{}^a(x)\,,\qquad \delta_\varepsilon \mathrm{T}{}^a{}_\mu(x)=-igf^{abc}\mathrm{T}{}^b{}_\mu(x)\varepsilon^c(x)\,.
\end{equation}
This relation follows from the covariant structure of all the involved quantities. The gauge transformation parameters are functions of space-time coordinates $\varepsilon^a(x), a=1,\ldots,n$. The gauge symmetry is non-Abelian,
\begin{equation}
[\delta_{\varepsilon_1},\delta_{\varepsilon_2}]=\delta_{[\varepsilon_1,\varepsilon_2]}\,,\qquad [\varepsilon_1,\varepsilon_2]{}^a=igf{}^{abc}\varepsilon_1{}^b\varepsilon_2{}^c\,.
\end{equation}
The interacting theory (\ref{EoM}) does not follow from the least action principle. So, the gauge identity in this model is not an immediate consequence of gauge invariance \cite{KLS13}.  The gauge 
identity is found in the following form:
\begin{equation}\label{Nid}
\mathcal{D}{}^\mu T{}{}_\mu=0\,,\qquad (\mathcal{D}_\mu){}^{ab}=(D_\mu){}^{ab}-ig\gamma f{}^{abc}X{}^c{}_\mu\,,
\end{equation}
where $\mathcal{D}$ is the prolonged covariant derivative. The proof of the gauge identity uses the definition of the covariant derivative (\ref{D-cov}) and relations (\ref{Jid}), (\ref{FG-tr}). The gauge symmetry and gauge identity are generated by different operators. In the Lagrangian theories, the gauge symmetries and gauge identities are generated by one and the same generator (so-called Second Noether theorem).  So, equations (\ref{EoM}) do not follow from the least action principle.

The concept of consistency of interaction in the class of non-Lagragian theories is explained in the article \cite{KLS13}. The interaction is termed consistent it preserves numbers of 
gauge symmetries, gauge identities, and physical degrees of freedom.  As is seen from (\ref{EoM}), (\ref{gt}), (\ref{Nid}), the interaction preserves all the gauge symmetries and gauge identities of free model. The number of physical degrees of freedom of not necessarily Lagrangian model is determined by the covariant formula (8) from \cite{KLS13}. It is based on the estimates of orders of gauge symmetries, gauge identities, and equations of motion. The interaction preserves the number of physical degrees of freedom because these orders do not change after addition of non-linear terms. All these means that equations (\ref{EoM}) describe consistent couplings between the multiplet of gauge vector fields.  At the free level, the vector fields are subjected to extended Chern-Simons equations.

\section{Stability}

The paper \cite{KLS14} tells us that the classical stability of the dynamics can be ensured by the existence of 
a bounded conserved quantity. In the class of field theories, these quantity is usually associated with the 00-component of 
second-rank conserved tensor.

The model (\ref{EoM}) admits a second-rank conserved tensor,
\begin{equation}\label{Theta-C}\begin{array}{c}\displaystyle
	\Theta^{\mu\nu}=G{}^a{}^\mu G{}^a{}^\nu-\frac12\eta^{\mu\nu}G{}^a{}_\rho G{}^a{}^\rho+
	\beta(G{}^a{}^\mu F{}^a{}^\nu+G{}^a{}^\nu F{}^a{}^\mu-\eta^{\mu\nu}G{}^a{}_\rho F{}^a{}^\rho)+\\[5mm]\displaystyle
	+(\beta\alpha_2-\alpha_1)(G{}^a{}^\mu G{}^a{}^\nu-\frac12\eta^{\mu\nu}G{}^a{}_\rho G{}^a{}^\rho)\,.
\end{array}\end{equation}
Expression (\ref{Theta-C}) is given by a covariantization of a selected conserved tensor of free extended Chern-Simons model \cite{KKL15}.  The value of
$\beta$ determines a particular conserved tensor that is preserved at interacting level. The divergence of the conserved tensor (\ref{Theta-C}) reads
\begin{equation}\label{Theta}\begin{array}{c}\displaystyle
	\partial_\mu\Theta^{\mu\nu}=-\varepsilon^{\nu\mu\rho}X^{a}{}_\mu\mathrm{T}{}^a{}_\rho\,.
\end{array}\end{equation}
 The 00-component of the tensor (\ref{Theta-C}) reads
\begin{equation}\label{Theta-00}\begin{array}{c}\displaystyle
	\Theta^{00}=\frac{1}{2}(G{}^a{}^0 G{}^0{}^0+G{}^a{}^i G{}^a{}^i)+
	\beta(G{}^a{}^0 F{}^a{}^0+G{}^a{}^i F{}^a{}^i)+\frac{1}{2}(\beta\alpha_2-\alpha_1)(G{}^a{}^0 G{}^a{}^0+G{}^a{}^i G{}^a{}^i)\,.
\end{array}\end{equation}
This relation follows from the transversality conditions (\ref{FG-tr}), Jacobi identity (\ref{Jid}), and antisymmetry property of 
structure constants $f^{abc}$. The quantity (\ref{Theta-00}) is nonnegative if it 
is given by a positive definite quadratic form of arguments $G{}^a{}^\mu,F{}^a{}^\mu$. The necessary and sufficient conditions 
are given in equation (25) of \cite{AKL19}. In the case at hands, it reads
\begin{equation}\label{g00}
	\gamma>0\,.
\end{equation}
So, the 00-component of the second-rank tensor (\ref{Theta-C}) can be bounded. The condition (\ref{g00}) selects the stable couplings in 
the class of interacting models (\ref{EoM}). All this means that the equations (\ref{EoM}), (\ref{g00}) describe the stable higher-derivative theory with the 
non-Abelian Yang-Mills gauge symmetry. To my knowledge this is a first model of such a kind.

\end{document}